\documentclass{article}

\usepackage{hyperref}
\usepackage[utf8]{inputenc}
\usepackage{latexsym,graphicx,amssymb,amsmath,mathrsfs,bbold} 

\newcommand{\bea}{\begin{eqnarray}}
\newcommand{\eea}{\end{eqnarray}}
\newcommand{\be}{\begin{equation}}
\newcommand{\ee}{\end{equation}}
\newcommand{\Tr}{\,\textrm{Tr}\,}
\newcommand{\GeV}{\,\textrm{GeV}\,}
\newcommand{\MeV}{\,\textrm{MeV}\,}
\newcommand{\crit}{\,\textrm{crit}\,}
\newcommand{\ns}{\,\textrm{ns}\,}
\newcommand{\s}{\,\textrm{s}\,}
\newcommand{\inter}{\,\textrm{int}\,}
\newcommand{\fm}{\,\textrm{fm}\,}

\newcommand{\liq}{\,\textrm{liq}\,}

\begin{document} 

\title{Perturbative RG analysis of the condensate dependence of the axial anomaly in the three flavor linear sigma model} 
\author{G. Fej\H{o}s \\
Institute of Physics, E\"otv\"os University \\
Budapest, H-1117, Hungary\\
e-mail: gergely.fejos@ttk.elte.hu}
\date{}

\maketitle

\begin{abstract}
{Coupling of `t Hooft's determinant term is investigated in the framework of the three flavor linear sigma model as a function of the chiral condensate. Using perturbation theory around the minimum point of the effective action, we calculate the renormalization group flow of the first field dependent correction to the coupling of the conventional $U_A(1)$ breaking determinant term. It is found that at low temperatures mesonic fluctuations make the anomaly increase when the chiral condensate decreases. As an application, we analyze the effect at the zero temperature nuclear liquid--gas transition.}
\end{abstract}

\section{Introduction}

The $U_A(1)$ subgroup of approximate $U_L(3)\times U_R(3)$ chiral symmetry is anomalously broken in quantum chromodynamics (QCD). QCD is a strongly coupled theory, and as such, most accurate results can be expected to emerge through lattice simulations. These, however, lack the ability to simulate the system at finite density due to the notorius sign problem. To tackle this issue, it is common to build effective models upon chiral symmetry, which are expected to capture essential features of QCD in the low energy regime. Even at zero density, these models are known to provide reasonable results for temperatures below that of the chiral transition \cite{fukushima10}.

In effective theories, such as the Nambu--Jona--Lasinio or linear sigma models this is taken into account by `t Hooft's determinant term. Coefficients of operators  in the Lagrangian of field theories, including the aforementioned determinant term, are considered to be (coupling) constants, without any field or environment dependence. In the quantum version of the action, however, fluctuations introduce temperature ($T$), baryochemical potential ($\mu_B$) and also field dependence as they become coefficient {\it functions}. When talking about field dependence of a given coupling one has in mind the resummation of higher dimensional operators that  can reappear when Taylor expanding the coefficient functions in terms of the field variable(s) around a conveniently chosen expansion point. 

In QCD, it is well established that the anomalous breaking of $U_A(1)$ symmetry should gradually disappear beyond the critical temperature, as at high $T$ the instanton density causing the anomaly exponentially vanishes \cite{thooft76,gross81}. At lower temperatures, however, the situation is far from being understood in a satisfactory fashion. One has also great interest in gaining results regarding the anomaly evolution at finite $\mu_B$ due to the sign problem, as mentioned earlier.

The finite temperature and/or density behavior of the $U_A(1)$ anomaly represents an active direction of research. More conservative results usually argue that the evaporation of the anomaly should follow that of the chiral condensate and thus the $U_A(1)$ symmetry restores around the critical temperature ($T_C$)  of the chiral transition\cite{mitter14,cossu15,ishii16,gomez-nicola18,bottaro20}. There are also several arguments and results that indicate that it is visible even beyond $T_C$ \cite{dick15,sharma16,fejos16,li20,braun20}. For example, earlier renormalization group studies indicate that when considering a field dependent anomaly coefficient, it decreases as a function of the chiral condensate \cite{fejos16,fejos18}, and this profile function can also depend explicitly on the tempereture (though the former effect is more dominant). Effective restoration of the anomaly has, e.g., a consequence regarding the order of the chiral transition \cite{pelissetto13}, the axion mass \cite{horvatic19}, and the fate of $\eta'$ meson, whose mass if substantially drops \cite{nagahiro06,sakai13,sakai17,jido19,horvatic19b,rai20}, in a nuclear medium could lead to the formation of an $\eta'$-nucleon bound state.

The goal of this paper is to calculate the first field dependent correction to the coupling of the `t Hooft determinant in the effective action perturbatively, i.e., we determine the first Taylor coefficient of the anomaly function. Determining this coefficient (at least qualitatively) allows for obtaining the behavior of the anomaly strength as the function of the chiral condensate. This allows for providing answer for the question whether the aforementioned results on anomaly strengthening \cite{fejos16,fejos18} can be reproduced within a simple perturbative renormalizatrion group setting of the linear sigma model, or determining a full functional dependence of the effective action is necessary. Fluctuations will be included using the functional variant of the renormalization group (FRG) \cite{kopietz10}, in the so-called local potential approximation (LPA). Even though renormalizable, we think of the model as an effective field theory, therefore, an ultra violet (UV) cutoff is inherently part of the system, which we set to $\Lambda = 1 \GeV$ (we expect the linear sigma model to emerge from QCD around this scale). Our task is to integrate out all fluctuations below $\Lambda$.

The paper is organized as follows. In Section 2, we introduce the model and the corresponding method of the FRG. Section 3 is devoted for calculating the effective action and discussing the problem of the expansion point of the Taylor series. After appropriate parametrization of the model, in Section 4, as an application, we show how the anomaly strengthens at the zero temperature nuclear liquid--gas phase transition. Section 5 contains the summary.

\section{Model and method}

The model we are working with in this paper is the three flavor linear sigma model, which is defined via the following Euclidean Lagrangian:
\bea
\label{Eq:Lag}
{\cal L}&=&\Tr(\partial_i M^\dagger \partial_i M) + m^2 \Tr (M^\dagger M) + g_1 \left(\Tr(M^\dagger M)\right)^2 + g_2\Tr (M^\dagger M  - \Tr (M^\dagger M)\cdot\mathbb{1})^2\nonumber\\
&&+a ( \det M + \det M^\dagger) - (h_0s_0+h_8s_8),
\eea
where $M$ contains the meson fields, $M=(s_a+i\pi_a)T_a$ [$T_a=\lambda_a/2$ are generators of the $U(3)$ group with $\lambda_a$ being the Gell-Mann matrices, $a=0,...,8$], $m^2$ is the mass parameter and $g_1$, $g_2$ refer to independent quartic couplings. As discussed in the previous section, the determinant term and the corresponding $a$ parameter is responsible for the $U_A(1)$ anomaly. We also have explicit symmetry breaking terms containing $h_0$ and $h_8$, which represent finite quark masses. 

Our main goal is to calculate the quantum effective action, $\Gamma$, built upon the theory defined via (\ref{Eq:Lag}). As announced in the introduction, we think of (\ref{Eq:Lag}) as an inherently effective model, which is only valid up to the scale $\Lambda = 1 \GeV$, therefore, one needs to take into account fluctuations with a cutoff $\Lambda$. The scale dependent quantum effective action, $\Gamma_k$, which includes fluctuations with momenta larger than $k$ (i.e., they are integrated out) is defined as
\bea
Z_k[J]&=&\int {\cal D} M {\cal D}M^\dagger \exp \Big\{-\int {\cal L} - \int (J M + h.c.) - \int \int M^\dagger R_k M \Big\}, \nonumber\\
\Gamma_k[M]&=&-\log Z_k[J]-\int (J M + h.c.) - \int \int M^\dagger R_k M,
\eea
where we omitted matrix indices, $J$ is the conjugate source variable to $M$, and $R_k$ is an appropriately chosen (bilocal) regulator function freezing fluctuations with momenta smaller than $k$. We note that the integrals can be considered either in direct or Fourier spaces.
The $\Gamma_k$ functional, for homogeneous field configurations, obeys the following, so-called flow equation \cite{kopietz10}:
\bea
\label{Eq:floweq}
\partial_k \Gamma_k [M] = \frac12 \Tr  \int_x \int_{q} (\Gamma_k^{(2)}+R_k)^{-1}(q)\partial_k R_k (q),
\eea 
where $\Gamma_k^{(2)}$ is the second functional derivative of $\Gamma_k$ in a homogeneous background field $M$, thus the $x$ integral merely gives a spacetime volume factor. We also assumed that the regulator is diagonal in momentum space.

Our aim is to calculate the scale dependent effective action, $\Gamma_k$, in an approximation that takes into account the evolution of the anomaly at the next-to-leading order, i.e., we wish to determine in $\Gamma_k$ the coefficient of the operator $\Tr(M^\dagger M)\cdot (\det M+\det M^\dagger)$. Our ansatz for $\Gamma_k$ is as follows:
\bea
\label{Eq:ansatz}
{\Gamma}_k&=&\int_x \Big[\Tr(\partial_i M^\dagger \partial_i M) + m^2_k \Tr (M^\dagger M) + g_{1,k} \left(\Tr(M^\dagger M)\right)^2 + g_{2,k}\Tr (M^\dagger M  - \Tr (M^\dagger M)\cdot\mathbb{1})^2\nonumber\\
&&+a_k ( \det M + \det M^\dagger) + a_{1,k} \Tr(M^\dagger M)\cdot (\det M + \det M^\dagger)- (h_0s_0+h_8s_8)\Big].
\eea
This is sometimes called the Local Potential Approximation (LPA), where momentum dependence is only introduced into the two point function, via the standard kinetic term in ({\ref{Eq:ansatz}). Note that the LPA can be considered as the leading order of the derivative expansion, and there is substantial evidence that these kind of series do converge \cite{kopietz10}. As seen in (\ref{Eq:ansatz}), instead of working with a completely general field dependent potential, we are employing perturbation theory in terms of the small parameter $1/\Lambda$. That is, by gradually including higher dimensional operators, since their coefficients scale with inverse powers of the scale, the ansatz (\ref{Eq:ansatz}) in the UV can be thought of as a power series in $1/\Lambda$. We choose (\ref{Eq:ansatz}) to be compatible with (\ref{Eq:Lag}), but all couplings come with $k$--dependence. The only exceptions are $h_0$ and $h_8$, as one point couplings do not flow with respect to the scale. Furthermore, notice the new term proportional to $a_{1,k}$, which is key for our purposes to determine the anomaly behavior at $k=0$. First, our task is to calculate $\Gamma_k^{(2)}$ from (\ref{Eq:ansatz}), then plug it into (\ref{Eq:floweq}), and identify the individual differential equations for $m_k^2$, $g_{1,k}$, $g_{2,k}$, $a_k$, and $a_{1,k}$. Finally, these equations need to be integrated from $k=1 \GeV$ to $k=0$ to obtain $\Gamma\equiv \Gamma_{k=0}$. In the ansatz (\ref{Eq:ansatz}), obviously the actual strength of the anomaly is not described by the parameter $a$, but rather $a+a_1\cdot \Tr(M^\dagger M)|_{\min}$, where we need to evaluate the chiral condensates in the minimum point of the effective action. Therefore, what we are basically after is the relative sign of $a_1$ to $a$ at $k=0$ to decide whether the anomaly strengthens or weakens as the chiral condensate gradually evaporates.

We finally note that there are various choices for the regulator function, $R_k$. In this paper we will stick to $R_k(q,p)\equiv R_k(q)(2\pi)^3 \delta(q+p)=(k^2-q^2)\Theta(k^2-q^2)(2\pi)^4 \delta(q+p)$, where $\Theta(x)$ is the step function. This variant has been shown to be the optimal choice for the LPA \cite{litim00}, maximizing the radius of convergence of an amplitude expansion.

\section{Calculation of the effective action}

The first step is to calculate $\Gamma_k^{(2)}$. In principle it is a $18\times 18$ matrix in the $s^a-\pi^a$ space, and there is not much hope that one can invert such a complicated expression analytically. Luckily, it is not necessary at all, as in (\ref{Eq:ansatz}) we kept the field dependence up to the order of ${\cal O}(M^5)$. By working with a restricted background, $\Gamma_k^{(2)}$ is easily invertable and by expanding the rhs of (\ref{Eq:floweq}) in terms of the field variables it still allows for identifying each operator that are being kept in (\ref{Eq:ansatz}).

A convenient choice is to work with $M=s_0T_0+s_8T_8$. In such a background, the operators that need to be identified are as follows:
\begin{subequations}
\label{Eq:operators}
\bea
\rho&:=&\Tr (M^\dagger M)=\frac12 (s_0^2+s_8^2), \\
\tau&:=&\Tr (M^\dagger M - \Tr (M^\dagger M)\cdot\mathbb{1})^2=\frac{1}{24} s_8^2 (8 s_0^2 - 4 \sqrt{2} s_0 s_8 + s_8^2), \\
\Delta&:=&\det M+\det  M^\dagger=\frac{1}{36} (2 \sqrt{6} s_0^3 - 3 \sqrt{6} s_0 s_8^2 - 2 \sqrt{3} s_8^3).
\eea
\end{subequations}
The $\Gamma_k^{(2)}$ matrix elements in the scalar sector read
\bea
\Gamma_{k,s_0s_0}^{(2)}&\!\!\!=\!\!\!&q^2+m_k^2+g_{1,k}(3s_0^2+s_8^2)+\frac23g_{2,k}s_8^2+\sqrt{\frac23}a_ks_0+a_{k,1}\Big(\frac53\sqrt{\frac23}s_0^3-\frac{1}{2\sqrt6}s_0s_8^2-\frac{1}{6\sqrt3}s_8^3\Big), \\
\Gamma_{k,s_0s_8}^{(2)}&\!\!\!=\!\!\!&2g_{1,k} s_0s_8+g_{2,k}\Big(\frac43s_0s_8-\frac{1}{\sqrt2}s_8^2\Big)-\frac{a_k}{\sqrt6}s_8-a_{1,k}\Big(\frac{1}{2\sqrt6}s_0^2s_8+\frac{1}{2\sqrt3}s_0s_8^2+\frac{1}{\sqrt6}s_8^3\Big),\\
\Gamma_{k,s_8s_8}^{(2)}&\!\!\!=\!\!\!&q^2+m_k^2+g_{1,k}(s_0^2+3s_8^2)+g_{2,k}\Big(\frac23s_0^2-\sqrt2 s_0s_8+\frac12s_8^2\Big)-a_k\Big(\frac{1}{\sqrt6}s_0+\frac{1}{\sqrt3}s_8\Big)\nonumber \\
&&-a_{1,k}\Big(\frac{1}{6\sqrt6}s_0^3+\frac{1}{2\sqrt3}s_0^2s_8+\sqrt{\frac32}s_0s_8^2+\frac{5}{3\sqrt3}s_8^3\Big),\\
\Gamma_{k,s_1s_1}^{(2)}&\!\!\!=\!\!\!&\Gamma_{k,s_2s_2}^{(2)}=\Gamma_{k,s_3s_3}^{(2)}=q^2+m_k^2+g_{1,k}(s_0^2+s_8^2)+g_{2,k}\Big(\frac23s_0^2+\sqrt2 s_0s_8+\frac16s_8^2\Big)\nonumber\\
&&\hspace{2.6cm}+a_k\Big(-\frac{1}{\sqrt6}s_0+\frac{1}{\sqrt3}s_8\Big)+a_{1,k}\Big(-\frac{1}{6\sqrt6}s_0^3+\frac{1}{2\sqrt3}s_0^2s_8-\frac{1}{\sqrt6}s_0s_8^2+\frac{1}{3\sqrt3}s_8^3\Big), \nonumber\\ 
\eea
\bea
\Gamma_{k,s_4s_4}^{(2)}&\!\!\!=\!\!\!&\Gamma_{k,s_5s_5}^{(2)}=\Gamma_{k,s_6s_6}^{(2)}=\Gamma_{k,s_7s_7}^{(2)}=q^2+m_k^2+g_{1,k}(s_0^2+s_8^2)+g_{2,k}\Big(\frac23s_0^2- \frac{1}{\sqrt2}s_0s_8+\frac16s_8^2\Big)\nonumber\\
&&\hspace{4.1cm}-a_k\Big(\frac{1}{\sqrt6}s_0+\frac{1}{2\sqrt3}s_8\Big)\nonumber\\
&&\hspace{4.1cm}-a_{1,k}\Big(\frac{1}{6\sqrt6}s_0^3+\frac{1}{4\sqrt3}s_0^2s_8+\frac{1}{\sqrt6}s_0s_8^2+\frac{5}{12\sqrt3}s_8^3\Big), 
\eea
while the pseudoscalar components are
\bea
\Gamma_{k,\pi_0\pi_0}^{(2)}&\!\!\!=\!\!\!&q^2+m_k^2+g_{1,k}(s_0^2+s_8^2)-\sqrt{\frac23}a_k s_0-a_{1,k}\Big(\frac13\sqrt{\frac{2}{3}}s_0^3+\frac12\sqrt{\frac{3}{2}}s_0s_8^2+\frac{1}{6\sqrt3}s_8^3\Big), \\
\Gamma_{k,\pi_0\pi_8}^{(2)}&\!\!\!=\!\!\!&g_{2,k}\Big(\frac23 s_0s_8-\frac{1}{3\sqrt2}s_8^2\Big)+\frac{a_k}{\sqrt6}s_8+a_{1,k}\Big(\frac{1}{2\sqrt6}s_0^2s_8+\frac{1}{2\sqrt6}s_8^3\Big), \\
\Gamma_{k,\pi_8\pi_8}^{(2)}&\!\!\!=\!\!\!&q^2+m_k^2+g_{1,k}(s_0^2+s_8^2)+g_{2,k}\Big(-\frac{\sqrt2}{3}s_0s_8+\frac16 s_8^2\Big)+a_k\Big(\frac{1}{\sqrt6}s_0+\frac{1}{\sqrt3}s_8\Big)\nonumber\\
&&+a_{1,k}\Big(\frac{5}{6}\sqrt{\frac16}s_0^3+\frac{1}{2\sqrt3}s_0^2s_8+\frac{1}{3\sqrt3}s_8^3\Big), \\
\Gamma_{k,\pi_1\pi_1}^{(2)}&\!\!\!=\!\!\!&\Gamma_{k,\pi_2\pi_2}^{(2)}=\Gamma_{k,\pi_3\pi_3}^{(2)}=q^2+m_k^2+g_{1,k}(s_0^2+s_8^2)+g_{2,k}\Big(\frac{\sqrt2}{3}s_0s_8-\frac16 s_8^2\Big)\nonumber\\
&&\hspace{2.8cm}+a_k\Big(\frac{1}{\sqrt6}s_0-\frac{1}{\sqrt3}s_8\Big)+a_{1,k}\Big(\frac{5}{6}\sqrt{\frac16}s_0^3-\frac{1}{2\sqrt3}s_0^2s_8-\frac{2}{3\sqrt3}s_8^3\Big), \\
\Gamma_{k,\pi_4\pi_4}^{(2)}&\!\!\!=\!\!\!&\Gamma_{k,\pi_5\pi_5}^{(2)}=\Gamma_{k,\pi_6\pi_6}^{(2)}=\Gamma_{k,\pi_7\pi_7}^{(2)}=q^2+m_k^2+g_{1,k}(s_0^2+s_8^2)+g_{2,k}\Big(-\frac{1}{3\sqrt2}s_0s_8+\frac56 s_8^2\Big)\nonumber\\
&&\hspace{4.3cm}+a_k\Big(\frac{1}{\sqrt6}s_0+\frac{1}{2\sqrt3}s_8\Big)+a_{1,k}\Big(\frac{5}{6}\sqrt{\frac16}s_0^3+\frac{1}{4\sqrt3}s_0^2s_8+\frac{1}{12\sqrt3}s_8^3\Big). \nonumber\\
\eea
Using that $\partial_k R_k(q)=2k\Theta(k^2-q^2)$ and $\Gamma^{(2)}_k(q)+R_k(q)=\Gamma^{(2)}_k(k)$ for $q<k$, from (\ref{Eq:floweq}) we get
\bea
\label{Eq:floweq2}
\partial_k \Gamma_k = \int_x \frac{k^5}{32\pi^2} \Tr \big(\Gamma_k^{(2)}(k)\big)^{-1}.
\eea
Plugging in the matrix elements calculated in the $M=s_0T_0+s_8T_8$ background, we can expand the rhs of (\ref{Eq:floweq2}) in terms of $s_0$ and $s_8$. After this step, using (\ref{Eq:operators}) we identify the $\rho$, $\tau$ and $\Delta$ operators as
\bea
\label{Eq:floweqansatz}
\partial_k \Gamma_k = \int_x \Big[ \partial_k m_k^2 \cdot \rho + \partial_k g_{1,k} \cdot \rho^2 + \partial_k g_{2,k} \cdot \tau + \partial_k a_k \cdot \Delta + \partial_k a_{1,k} \cdot \rho \Delta+...\Big],
\eea
where
\bea
\label{Eq:flowcoup0}
\partial_k m_k^2&\!\!\!=\!\!\!& - \frac{k^5}{32\pi^2} \frac{8(15g_{1,k}+4g_{2,k})}{3(k^2+m_k^2)^2}, \quad \partial_k g_{1,k} = \frac{k^5}{32\pi^2}  \frac{8(117g_{1,k}^2+48g_{1,k}g_{2,k}+16g_{2,k}^2)}{9(k^2+m_k^2)^3}, \nonumber\\
\partial_k g_{2,k} &\!\!\!=\!\!\!& \frac{k^5}{32\pi^2} \frac{48g_{1,k}g_{2,k}+32g_{2,k}^2}{(k^2+m_k^2)^3}, \quad \partial_k a_k = \frac{k^5}{32\pi^2}\Big(\frac{8a_k(3g_{1,k}-4g_{2,k})}{(k^2+m_k^2)^3}-\frac{24a_{1,k}}{(k^2+m_k^2)^2}\Big), \nonumber\\
\partial_k a_{1,k} &\!\!\!=\!\!\!& \frac{k^5}{32\pi^2}\bigg(\frac{32a_k(-9g_{1,k}^2+6g_{1,k}g_{2,k}+2g_{2,k}^2)}{(k^2+m_k^2)^4}+\frac{16a_{1,k}(33g_{1,k}-2g_{2,k})}{3(k^2+m_k^2)^3}\bigg).
\eea
Note that we treated the anomaly as perturbation and dropped every term beyond ${\cal O}(a_k,a_{1,k})$. Introducing scale independent variables, from (\ref{Eq:flowcoup0}) one easily reproduces the well known 1-loop $\beta$ functions of the couplings in the linear sigma model \cite{pisarski84}. Our task now is to solve Eqs. (\ref{Eq:flowcoup0}) starting from $k=\Lambda\equiv 1\GeV$ to $k=0$. 

Solving (\ref{Eq:flowcoup0}) all the way down to $k=0$ would require $m_k^2>0$ throughout the renormalization group flow. Since we wish to obtain phenomenologically reasonable results, the potential has to show spontaneous symmetry breaking. That is to say, when all fluctuations are integrated out, $m^2_{k=0}$ has to be negative. But then there exists a critical scale $k_{\crit}>0$, for which all denominators in (\ref{Eq:flowcoup0}) blow up and the flow equations lose their meaning. The way out is to realize is that one actually has the choice to determine the flow equations in the minimum point of the effective action, $\Gamma_k|_{s_{0,\min},s_{8,\min}}$, rather than evaluating it in a vanishing background. That is, all renormalization group flows are to be extracted at $s_{0,\min}$, $s_{8,\min}$. This way one always has a positive definite denominators and the flow equation is valid for any $k$. 

One, therefore, repeats the calculations starting from (\ref{Eq:floweq2}), but this time expands only in terms of $s_8$ so that the $\rho$ dependence of the parameters can be traced via $s_0$. A long but straightforward calculation leads once again to the possibility of identifying the invariants appear in (\ref{Eq:floweqansatz}), whose coefficients now read as
\bea
\partial_k m_k^2 &\!\!\!=\!\!\!& -\frac{k^5}{32\pi^2} \Bigg[ \frac{18g_{1,k}}{(k^2+m_k^2+2g_{1,k}\rho_0)^2}+\frac{6g_{1,k}}{(k^2+m_k^2+6g_{1,k}\rho_0)^2}+\frac{16(3g_{1,k}+2g_{2,k})}{3(k^2+m_k^2+2g_{1,k}\rho_0+4g_{2,k}\rho_0/3)^2}\nonumber\\
&&\hspace{1cm}+\frac{72g^2_{1,k}\rho_0}{(k^2+m_k^2+2g_{1,k}\rho_0)^3}
+\frac{72g^2_{1,k}\rho_0}{(k^2+m_k^2+6g_{1,k}\rho_0)^3}+\frac{64(3g_{1,k}+2g_{2,k})^2\rho_0}{9(k^2+m_k^2+2g_{1,k}\rho_0+4g_{2,k}\rho_0/3)^3} \Bigg], \nonumber\\
\partial_k g_{1,k}&\!\!\!=\!\!\!&\frac{k^5}{32\pi^2}\Bigg[ \frac{36g_{1,k}^2}{(k^2+m_k^2+2g_{1,k}\rho_0)^3}+\frac{36g_{1,k}^2}{(k^2+m_k^2+6g_{1,k}\rho_0)^3}+\frac{32(3g_{1,k}+2g_{2,k})^2}{9(k^2+m_k^2+2g_{1,k}\rho_0+4g_{2,k}\rho_0/3)^3}\Bigg], \nonumber\\
\partial_k g_{2,k}&\!\!\!=\!\!\!&\frac{k^5}{32\pi^2}\Bigg[ \frac{6g_{2,k}^2}{(k^2+m_k^2+2g_{1,k}\rho_0)^3}-\frac{9g_{2,k}/2}{\rho_0(k^2+m_k^2+2g_{1,k}\rho_0)^2}+\frac{3g_{2,k}(6g_{1,k}+g_{2,k})}{\rho_0(g_{2,k}-3g_{1,k})(k^2+m_k^2+6g_{1,k}\rho_0)^2}\nonumber\\
&&\hspace{1cm}\frac{30g_{2,k}^2}{(k^2+m_k^2+2g_{1,k}\rho_0+4g_{2,k}\rho_0/3)^3}+\frac{3g_{2,k}(g_{2,k}-21g_{1,k})/2}{\rho_0(g_{2,k}-3g_{1,k})(k^2+m_k^2+2g_{1,k}\rho_0+4g_{2,k}\rho_0/3)^2} \Bigg],\nonumber\\
\partial_k a_k &\!\!\!=\!\!\!&\frac{k^5}{32\pi^2} \Bigg[-\frac{36g_{1,k}(a_k+2a_{1,k}\rho_0)}{(k^2+m_k^2+2g_{1,k}\rho_0)^3}-\frac{18(a_k+a_{1,k}\rho_0)}{\rho_0(k^2+m_k^2+2g_{1,k}\rho_0)^2}-\frac{12g_{1,k}(3a_k+10a_{1,k}\rho_0)}{(k^2+m_k^2+6g_{1,k}\rho_0)^3}\nonumber\\
&&\hspace{1cm}-\frac{6a_k+10a_{1,k}\rho_0}{\rho_0(k^2+m_k^2+6g_{1,k}\rho_0)^2}+\frac{16(3g_{1,k}+2g_{2,k})(3a_k+a_{1,k}\rho_0)}{3(k^2+m_k^2+2g_{1,k}\rho_0+4g_{2,k}\rho_0/3)^3}\nonumber\\
&&\hspace{1cm}+\frac{4(6a_k+a_{1,k}\rho_0)}{\rho_0(k^2+m_k^2+2g_{1,k}\rho_0+4g_{2,k}\rho_0/3)^2}\Bigg], \nonumber
\eea
\bea
\label{Eq:flowcoup}
\partial_k a_{1,k} &\!\!\!=\!\!\!&\frac{k^5}{32\pi^2} \Bigg[\frac{36g_{1,k}(a_k+2a_{1,k}\rho_0)}{\rho_0(k^2+m_k^2+2g_{1,k}\rho_0)^3}+\frac{9a_k}{\rho_0^2(k^2+m_k^2+2g_{1,k}\rho_0)^2}+\frac{12g_{1,k}(3a_k+10a_{1,k}\rho_0)}{\rho_0(k^2+m_k^2+6g_{1,k}\rho_0)^3} \nonumber\\
&&\hspace{1cm}-\frac{16(3g_{1,k}+2g_{2,k})(3a_k+a_{1,k}\rho_0)}{3\rho_0(k^2+m_k^2+2g_{1,k}\rho_0+4g_{2,k}\rho_0/3)^3}-\frac{12a_k}{\rho_0^2(k^2+m_k^2+2g_{1,k}\rho_0+4g_{2,k}\rho_0/3)^2}\Bigg],
\eea
where we have denoted the expansion point by $\rho_0$, which is to be set to the value of $\rho$ corresponding to the minimum point of the effective action [note that $\rho=\Tr (M^\dagger M)/2$]. As a side remark, one easily checks that choosing $\rho_0=0$ (\ref{Eq:flowcoup}) would lead back to the earlier results, (\ref{Eq:flowcoup0}). Our task is to integrate the system of equations (\ref{Eq:flowcoup}) from $k=\Lambda\equiv 1 \GeV$ down to $k=0$ with the boundary conditions $m_\Lambda^2=m^2$, $g_{1,\Lambda}=g_1$, $g_{2,\Lambda}=g_2$, $a_\Lambda=a$, $a_{1,\Lambda}=0$, where $m^2, g_1, g_2, a$ are such constants that reproduce as accurately as possible the mesonic spectrum in the infrared. Here we used that at the UV scale the coefficient of the operator $\rho\Delta$ can be set to zero due to perturbative renormalizability. This might be questionable if the UV scale was not high enough, as being a dimension 5 operator, dimensional analyis suggests that its coefficient, $a_1$, is of ${\cal O}(1/\Lambda)$. Obviously if the linear sigma model was not an effective theory, and $\Lambda$ could be sent to infinity, the term in question would not be present. But, in principle the $a_1$ coupling could be included already in the UV action. Investigation of such a scenario is beyond the scope of this paper.

Before solving the coupled system of equations (\ref{Eq:flowcoup}), we need to fix the explicit symmetry breaking terms, i.e., the values for $h_0$, $h_8$. Instead of $h_0$ and $h_8$, we will work in the nonstrange--strange basis, i.e., $h_{\ns}=\sqrt{\frac23}h_0+\frac{1}{\sqrt3}h_8$, $h_s=\frac{1}{\sqrt3}h_0-\sqrt{\frac{2}{3}}h_8$. The partially conserved axialvector current (PCAC) relations give
\bea
m_{\pi}^2f_{\pi}=h_{\ns}, \quad m_K^2f_K=\frac{h_{\ns}}{2}+\frac{h_s}{\sqrt2},
\eea
where $m_{\pi}^2=\delta^2 \Gamma/\delta \pi_i^2(q=0)$ $[i=1,2,3]$ and $m_K^2=\delta^2 \Gamma/\delta \pi_i^2(q=0)$ $[j=4,5,6,7]$. Using physical pion and kaon masses, $\sim 140 \MeV$, $\sim 494 \MeV$, respectively, and decay constants, $f_{\pi}=93 \MeV$, $f_K=113 \MeV$, one gets
\bea
\label{Eq:h}
h_{\ns}=m_{\pi}^2f_{\pi} \approx (122 \MeV)^3, \quad h_{\s}=\frac{1}{\sqrt2}(2m_K^2f_K-m^2_{\pi}f_{\pi}) \approx (335 \MeV)^3,
\eea
which is equivalent to
\bea
h_0&=&\sqrt{\frac23}\big(m_{\pi}^2f_{\pi}/2+m_K^2f_K)\approx (285 \MeV)^3, \quad h_8=\frac{2}{\sqrt3}\big(m_{\pi}^2f_{\pi}-m_K^2f_K) \approx -(310 \MeV)^3.
\eea
Now we use that Ward identities of chiral symmetry lead to
\bea
\frac{\delta \Gamma}{\delta s_{\ns}}(q=0)=m_{\pi}^2s_{\ns}-h_{\ns}, \quad \frac{\delta \Gamma}{\delta s_{\s}}(q=0)=\frac{m_K^2-m_{\pi}^2}{\sqrt2}s_{\ns}+m_K^2s_{\s}-h_{\s}.
\eea
Combined with (\ref{Eq:h}), this shows that no matter how we choose the remaining parameters $m^2, g_1, g_2, a$, in the minimum point of the effective action
\bea
s_{\ns,\min}=f_{\pi}, \quad s_{\s,\min}=\sqrt2(f_K-f_{\pi}/2).
\eea
That leads to $\rho_0=(s_{\ns,\min}^2+s_{\s,\min}^2)/2$, and thus we are ready to fix the aforementioned parameters. Solving (\ref{Eq:flowcoup}),  the values $\{m^2, g_1, g_2, a\}\approx \{0.6835 \GeV^2, 29.7, 91.5, -4.4 \GeV\}$ lead to the masses of the pion, kaon, $\eta$, $\eta'$ as $m_{\pi} \approx 133 \MeV$, $m_K \approx 494 \MeV$, $m_{\eta} \approx 537 \MeV$, $m_{\eta'}\approx 957 \MeV$, respectively. Note that the strength of the axial anomaly, $a$, is negative, and it remains so throughout the renormalization group flow. However, after solving (\ref{Eq:flowcoup}), one concludes that at $k=0$ the coefficient $a_{1,k=0}$ is positive. That is to say, since the actual strength of the determinant term is $A:=a_{k=0}+a_{1,k=0}\cdot \rho_0$, when the chiral condensate evaporates, the absolute value of $A$ becomes larger. This shows that at low temperatures $T$, where the $T$ dependence of the anomaly parameters is negligible, the anomaly is actually strengthening as chiral symmetry gradually restores. That is one of the main results of the paper.

In what follows, we provide a rough estimate how the anomaly behaves at the zero temperature nuclear liquid--gas transition.

\section{Anomaly strengthening at the nuclear liquid--gas transition}

In this section we apply our results to the zero temperature nuclear liquid--gas transition. We assume that the nucleon field couples to the mesons via Yukawa interaction, ${\cal L}_{\inter}=g\bar{\psi}M_5\psi$, $\psi^T=(p, n)$, $M_5 =\sum_{a=\ns,1,2,3} (s^a+i\pi^a\gamma_5)T^a$, where the nonstrange generator is $T^{\ns}=\sqrt{2/3}T^0+1/\sqrt3 T^8$, while $\gamma_5$ is the fifth Dirac matrix. In principle one would also need to include the dynamics of an $\omega$ vector particle into the system \cite{floerchinger12,drews16} that models the repulsive interaction between nucleons, but as we will see in a moment, for our purposes it plays no role.

First, we exploit some of the zero temperature properties of nuclear matter. Note that, in the current model, the nucleon mass entirely originates from the spontaneous breaking of chiral symmetry,
\bea
m_{N}(s_{\ns})=g_Ys_{\ns}/2,
\eea
and since $m_{N}(f_{\pi})\approx 939 \MeV$ in the vacuum, we arrive at $g_Y\approx 20.19$. Normal nuclear density, $n_N \approx 0.17 \fm^{-3} \approx (109.131 \MeV)^3$ leads to the Fermi momentum, $p_F$, of the nucleons, since at the mean field level, for $T=0$ we have
\bea
n_N=4\int_p n_F\Big(\sqrt{p^2+m_N^2}-p_F\Big)\Big|_{T=0}\equiv \frac{2}{3\pi^2}p_F^3,
\eea
therefore, $p_F \approx 267.9 \MeV \approx 1.36 \fm^{-1}$. This leads to the nonstrange condensate in the liquid phase, $s_{\ns\!,\liq}$, because the Landau mass, which is defined as
\bea
M_L=\sqrt{p_F^2+m_N^2(s_{\ns\!,\liq})}
\eea
is known to be $M_L \approx 0.8m_N(f_{\pi}) \approx 751.2 \MeV$, and thus $s_{\ns\!,\!\liq} \approx 69.52 \MeV$ \cite{floerchinger12,drews16}. This shows that as we increase the chemical potential, the nonstrange chiral condensate, $s_{\ns}$, jumps: $f_{\pi}\rightarrow s_{\ns\!,\liq}$. This will definitely be accompanied by a jump in the strange condensate, but it has been shown to be significantly smaller \cite{fejos18}. Neglecting the change in $s_{\s}$, the $\rho$ chiral invariant jumps as $(f_\pi^2+s_{\s\!,\min}^2)/2 \rightarrow (s_{\ns\!,\liq}^2+s_{\s,\min}^2)/2$. As discussed in the previous section, the anomaly strength is $A=a_{k=0}+a_{1,k=0}\cdot \rho$, which also jumps accordingly, and the change in $A$ becomes
\bea
\Delta A = a_{1,k=0}\cdot\Delta \rho,
\eea
where $\Delta \rho=(s_{\ns,\liq}^2-f_{\pi}^2)/2$. Solving (\ref{Eq:flowcoup}) one gets $a_{k=0}\approx -9.05\GeV$ and $a_{1,k=0} \approx 494.5\GeV^{-1}$, therefore, the relative change in the anomaly at the liquid--gas transition is
\bea
\frac{\Delta A}{A}=\frac{a_{1,k=0}\cdot\Delta \rho}{a_{k=0}+a_{1,k=0}\cdot\rho_0} \approx 0.2 = 20\%,
\eea
which is in the ballpark of the result of \cite{fejos18}. One can now check how robust this result is with respect to changing the cutoff $\Lambda$. A thorough investigation reveals that in a cutoff interval of $0.8$--$1.5 \GeV$, while the mass spectrum can be maintained within a few percent error after reparametrization, the $\Delta A/A$ ratio is less stable. One finds that the latter is is a monotonically decreasing function of the cutoff and varies roughly between $15$--$40\%$ in the above interval. Results show that when going below $1  \GeV$ the cutoff dependence gets stronger, which is understandable, since non-renormalizable operators are absent at the UV scale. That is, if the latter is chosen to be too small, the model cannot provide robust results (more operators would be needed). Going beyond $1.5$--$2$ GeV, in turn, would be physically inappropriate as at those scales quark degrees of freedom would definitely play a crucial role. From these findings it is safe to say is that the relative change of the anomaly strength is of ${\cal O}( 10\%)$ at the transition point.

At this point we once again wish to emphasize that we have neglected the drop in the strange condensate, and also, the present analysis is based on perturbation theory. In principle higher order operators that break the $U_A(1)$ subgroup should also be resummed, e.g., terms such as $\sim \big(\Tr (M^\dagger M)\big)^n (\det M+ \det M^\dagger)$ could be of huge importance. The lesson we wish to point out here is that the present, rather simple perturbative calculation can also capture the phenomenon of strengthening anomaly as the chiral condensate (partially) evaporates.

\section{Conclusions}

In this paper we investigated how the $U_A(1)$ anomaly behaves as a function of the chiral condensate. We worked with the three flavor linear sigma model, and calculated the leading correction in a $1/\Lambda$ expansion to the conventional anomaly term caused by quantum fluctuations. We have found that the coefficient of the aforementioned operator, $\sim \Tr (M^\dagger M)\cdot (\det M + \det M^\dagger)$, causes the actual strength of the anomaly to become larger once the chiral condensate evaporates. For the sake of an example, we demonstrated that at the zero temperature nuclear liquid--gas transition, where (on top of a jump in the nuclear density) the chiral condensate partially restores, the actual strength of the anomaly increases. This could also happen toward the full restoration of chiral symmetry, where quark dynamics also play a significant role. Note that our findings are based solely on calculating mesonic fluctuations, and no instanton effects have been taken into account.

The linear sigma model, being an effective field theory, cannot accommodate instantons as the fundamental model of QCD. Still, there are at least two directions worth exploring in the effective model framework. Recently it has been shown \cite{pisarski20} that $3Q$-point interactions are generated by instantons with $Q$ topological charge, which can be embedded into the linear sigma model via $\sim [(\det M^\dagger)^Q + (\det M)^Q]$ operators. Another important direction could be to assign environment dependence even to the bare anomaly coefficient(s) from QCD data and see how these compete against thermal effects caused by mesonic fluctuations.

Finally, we wish to point out that our study calls for an extension via a non-perturbative treatment, where fluctuations are taken into account beyond the ${\cal O}(a)$ order, and the coefficient function of the determinant term is obtained in a functional fashion, rather than at the lowest order of its Taylor series. The aforementioned directions are under progress and will be reported in a separate study.

\section*{Acknowledgements}

The author thanks A. Patk\'os and Zs. Sz\'ep for discussions on related topics. This research was supported by the National Research, Development and Innovation Fund under Project No. PD127982, by the János Bolyai Research Scholarship of the Hungarian Academy of Sciences, and by the ÚNKP-20-5 New National Excellence Program of the Ministry for Innovation and Technology from the source of the National Research, Development and Innovation Fund.

\makeatletter
\@addtoreset{equation}{section}
\makeatother

\end{document}